\documentclass[aps,showpacs,twocolumn,twoside,amsmath,
amssymb,superscriptaddress,prc]{revtex4-1}
\usepackage{graphicx}
\usepackage{tikz}
\usepackage{epstopdf}
\usepackage[percent]{overpic}  
\usepackage{amsmath}
\usepackage{scrextend}  
\usepackage{braket}
\usepackage{bm}
\usepackage{xcolor,xspace}
\usepackage[ulem=normalem]{changes}
\usepackage{hyperref}
\hypersetup{colorlinks=True,urlcolor=blue,linkcolor=blue,citecolor=blue,filecolor=black}

\colorlet{Changes@Color}{red}  

\newcommand{\ba}{\begin{eqnarray}}
\newcommand{\ea}{\end{eqnarray}}
\newcommand{\bsub}{\begin{subequations}}
\newcommand{\esub}{\end{subequations}}

\begin{document}
\title{Persistent vibrational structure in $^{110-116}$Cd}
\author{N.~Gavrielov}\email{noam.gavrielov@yale.edu}
\affiliation{Center for Theoretical Physics,
Sloane Physics Laboratory, Yale University, New Haven,
Connecticut 06520-8120, USA}
\affiliation{Racah Institute of Physics, The Hebrew University, Jerusalem 91904, Israel}
\author{J.E.~Garc\'\i a-Ramos}\email{enrique.ramos@dfaie.uhu.es}
\affiliation{Department of Integrated Sciences and
  Center for Advanced Studies in Physics, Mathematics
  and Computation, University of Huelva, 21071
  Huelva, Spain}
\author{P.~Van Isacker}\email{isacker@ganil.fr}
\affiliation{Grand Acc\'el\'erateur National d'Ions Lourds, CEA/DRF-CNRS/IN2P3,
Bvd Henri Becquerel, BP 55027, F-14076 Caen, France}
\author{A.~Leviatan}\email{ami@phys.huji.ac.il}
\affiliation{Racah Institute of Physics, The Hebrew University, Jerusalem 91904, Israel}

\date{\today}  

\begin{abstract}
  The empirical spectra and $E2$ decay rates in
  $^{110,112,114,116}$Cd are shown to be consistent with
  a vibrational interpretation for low-lying normal
  states, coexisting
  with a single deformed $\gamma$-soft band of
  intruder states.
  The observed deviations from this paradigm show up
  in particular
  non-yrast states, which are properly described by
  a Hamiltonian with
  U(5) partial dynamical symmetry. The latter is
  characterized by a good (broken) symmetry in most
  (in selected) normal states, weakly
  coupled to intruder states.
\end{abstract}

\maketitle

The concepts of shapes and symmetries play a pivotal role
in  the quest for understanding the structure and simple
patterns in complex many-body systems.
A~notable example is found in atomic nuclei, where these
notions are instrumental for interpreting the
collective motion exhibited by
a multitude of protons and neutrons subject to the
strong interaction.
Based on earlier ideas of Bohr and Kalckar~\cite{Bohr36,Bohr37}
and on Rainwater's suggestion~\cite{Rainwater50} that nuclei may
be intrinsically deformed, a standard description of the nucleus
was proposed in terms of a quantum liquid drop, which can vibrate
and, if deformed, also rotate. This is commonly referred to as
the (Bohr-Mottelson) collective model of the
nucleus~\cite{Bohr52,Bohr53,Bohr75}.
Particular limits of the model provide insightful paradigms
for the dynamics of spherical, axially-deformed and
non-axial shapes. These geometric benchmarks correspond in 
the algebraic interacting boson model (IBM)
of Arima and Iachello~\cite{Iachello87} to solvable
limits, associated with dynamical symmetries.

Recent advances in high resolution spectroscopy of non-yrast
states~\cite{Garrett22},
impart valuable input for testing and challenging
the accepted paradigms of collective motion in nuclei.
The present work examines the collective model
hypothesis of quadrupole oscillations about a spherical
shape, in relation to the cadmium isotopes ($Z\!=\!48$).
The latter since long have been considered as textbook
examples of spherical-vibrator nuclei and U(5) dynamical
symmetry~\cite{Bohr75,Iachello87,Casten00,Heyde04,Rowe10}.
On the other hand,
detailed studies, using complementary spectroscopic
methods, have provided evidence for marked deviations
from such a structural
paradigm~\cite{Garrett07,Casten92,Kadi03,Garrett08,
  Garrett10,Garrett12}.
Two approaches have been proposed to address these
unexpected findings.
The first questions the spherical-vibrational
character of the $^{110,112}$Cd isotopes, replacing it with
multiple coexistence of states with different deformed
shapes in the same nucleus, a view qualitatively supported
by a beyond-mean-field (BMF) calculation with the
Gogny D1S
energy density functional~\cite{Garrett19,Garrett20}.
A second approach is based on the recognition that
the reported deviations
from a spherical-vibrator behavior
show up in selected states, while most states
retain their vibrational character.
In the terminology of symmetry, this
implies that the symmetry in question
is broken only in a subset of states, hence is
partial~\cite{Leviatan11}.
Such a U(5) partial dynamical symmetry (PDS) approach
was applied in Ref.~\cite{Leviatan18}
to describe the properties of $^{110}$Cd.

In this Letter, it is shown that the U(5)-PDS approach
of Ref.~\cite{Leviatan18} can be
extended to give a coherent description of
a~{\em series} of cadmium isotopes
with mass number $A$=110--116.
Their properties are analyzed based on a vibrational
interpretation
coupled to the presence of intruder states.
It is by now widely accepted that the Cd isotopes
exhibit shape coexistence in their low-energy
spectrum~\cite{Gneuss71,Fielding77,Kumpulainen92,HeydeWood11}.
However, unlike the multi-shape version of coexistence in
Refs.~\cite{Garrett19,Garrett20},
only two coexisting configurations with different shapes
are proposed here:
a spherical one, exhibiting an anharmonic vibrational
spectrum for normal states,
and one that is prolate deformed with $\gamma$-soft
characteristics,
that is, an axially-symmetric shape that can easily
turn triaxial, for intruder states.
The anharmonicity is due to the
presence of terms in the Hamiltonian
that break the U(5) symmetry in selected normal states,
and is essential
to reproduce the unexpected observed $E2$ decay patterns.
In this respect, it should be mentioned that
previous attempts to explain the experimental $E2$ matrix
elements relied on
strong mixing between spherical and intruder states
and ultimately proved
unsuccessful~\cite{Heyde82,Casten92,Deleze93,Kadi03,Garrett07,Garrett08,Garrett10}.
\begin{figure}[b]
\begin{overpic}[width=1\linewidth]{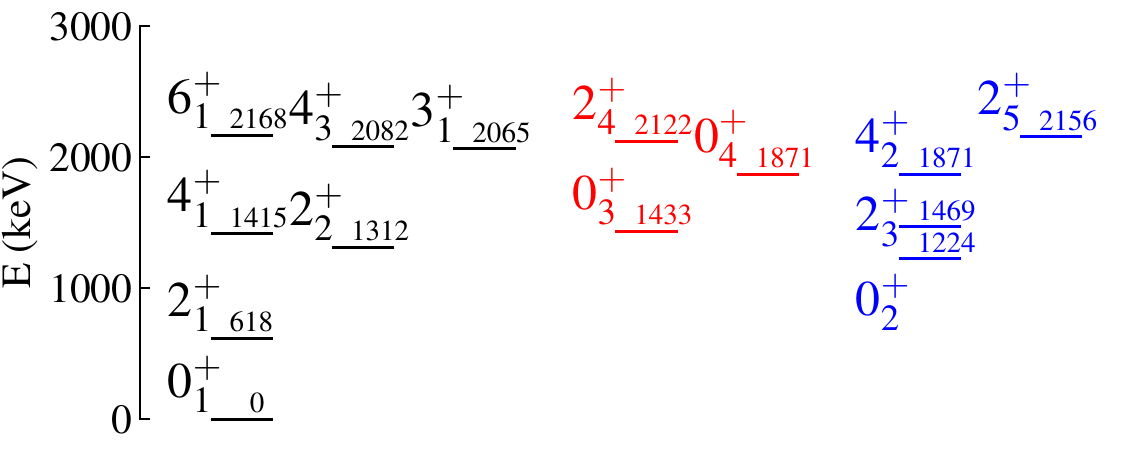}
\put (40,1) {\Large $^{112}$Cd}
\end{overpic}
\caption{
  Experimental energy
  levels
  in keV of $^{112}$Cd~\cite{ensdf}.
  Normal states are marked in black
  or in red if their $E2$ decays deviate from
  those of a spherical vibrator.
  Intruder states are marked in blue.
  \label{fig:112cd-schematic}}
\end{figure}
\begin{table*}
\begin{center}
  \caption{\label{Tab-1}
\small
Experimental (EXP) $B(E2)$ values in Weisskopf units (W.u.)
and quadrupole moments $Q(2^{+}_1)$ in eb,
for normal levels in $^{110-116}$Cd,
compared to calculated U(5)-DS and PDS values.
The $0^+_\alpha$ ($2^+_\alpha$) state corresponds
to the experimental $0^+_3,0^+_3,0^+_3,0^+_2$
($2^+_5,2^+_4,2^+_5,2^+_4$) state
for $^{A}$Cd ($A\!=\!110,112,114,116$), respectively.
In the U(5)-DS classification,
($0^{+}_1, 2^{+}_1, 2^{+}_2,4^{+}_1,6^{+}_1$) are class-A states
with $n_d\!=\!0,1,2,2,3$ and ($0^+_\alpha,2^+_\alpha$)
are states with $n_d\!=\!(2,3)$.
The U(5)-PDS calculations are
obtained using $\hat{T}(E2)$, Eq.~(\ref{T-E2}),
with parameters given in Fig.~\ref{fig:params}.
Data is taken from 
\cite{ensdf,Garrett12,Garrett07,Garrett08,Garrett10}.}
\begin{tabular}{ccccclccclccclccc}
\hline\noalign{\smallskip}
  &  &
\multicolumn{3}{c}{$^{110}$Cd} & &
\multicolumn{3}{c}{$^{112}$Cd} & &
\multicolumn{3}{c}{$^{114}$Cd} & &
\multicolumn{3}{c}{$^{116}$Cd} \\[1mm]
\cline{3-5}
\cline{7-9}
\cline{11-13}
\cline{15-17}\noalign{\smallskip}
$L_i$ $\;$  & $L_f$ $\;$ &
EXP  & U(5)-DS & PDS & &
EXP  & U(5)-DS & PDS & &
EXP  & U(5)-DS & PDS & &
EXP  & U(5)-DS & PDS \\ [1pt]
\noalign{\smallskip}\hline
$2^+_{1}$ & $0^+_{1}$ &
27.0(8)   & 27.0 & 27.0 & &
30.31(19) & 30.31 & 30.31 & &
31.1(19)  & 31.1 & 31.1 & &
33.5(12)  & 33.5 & 33.5 \\
$4^+_{1}$ & $2^+_{1}$ &
42(9)  & 46 & 46 & &
63(8)  & 53 & 52 & &
62(4)  & 55 & 55 & &
56(14) & 59 & 60 \\
$2^+_{2}$ & $2^+_{1}$ &
30(5)  & 46 & 45 & &
39(7)  & 53 & 51 & &
22(6)  & 55 & 53 & &
25(10) & 59 & 59 \\
$2^+_{2}$ & $0^+_{1}$ &
0.68(14) & 0 & 0.0 & &
0.65(11) & 0 & 0.0 & &
0.48(6)  & 0 & 0.0 & &
1.11(18) & 0 & 0.0 \\
$6^+_{1}$ & $4^+_{1}$ &
40(30)  & 58 & 53 & &
        & 68 & 59 & &
119(15) & 72 & 70 & &
110$^{+40}_{-80}$ & 75 & 79 \\
$0^+_{\alpha}$ & $2^+_{1}$ &
$<$ 7.9    & 46 & 0.08 & &
0.0121(17) & 53 & 0.01 & &
0.0026(4)  & 55 & 0.0026 & &
0.79(22) & 59 & 0.79 \\
$0^+_{\alpha}$ & $2^+_{2}$ &
$<$ 1680 & 0 & 43 & &
99(16)   & 0 & 49 & &
127(16)  & 0 & 61 & &
         & 0 & 60 \\        
$2^+_{\alpha}$ & $2^+_{2}$ &
0.7$^{+0.5}_{-0.6}$  & 11 & 0.124 & &
$<$ 1.6$^{+6}_{-4}$ & 13 & 0.08 & &
2.5$^{+16}_{-14}$    & 14 & 0.005 & &
2.0(6)            & 14 & 0.004 \\
$2^+_{\alpha}$  & $0^+_{\alpha}$ &
24.2(22) & 27 & 21 & &
25(7)    & 32 & 28 & &
17(5)    & 34 & 33 & &
35(10)   & 35 & 33 \\[2mm]
\hline\hline
$Q(2^{+}_1)$ & &
-0.40(3) & 0 & -0.13 & &
-0.38(3) & 0 & -0.13 & &
-0.35(5) & 0 & -0.13 & &
-0.42(4) & 0 & -0.14  \\
\noalign{\smallskip}\hline
\end{tabular}
\end{center}
\end{table*} 

Vibrations of spherical nuclei can be described in the
U(5) dynamical symmetry (DS) limit of the IBM,
associated with the chain, 
$\text{U(6)} \supset \text{U(5)} \supset
\text{SO(5)} \supset \text{SO(3)}$.
The DS basis states $\ket{[N],n_d,\tau,n_{\Delta},L}$
have quantum numbers which are the labels of
irreducible representations of the algebras in the chain. 
Here $N$ is the total number of monopole ($s$) and
quadrupole ($d$)  bosons, $n_d$ and $\tau$ are the
$d$-boson number and seniority, respectively,
$L$ is the angular momentum and $n_{\Delta}$ is a multiplicity
label. The U(5)-DS Hamiltonian can be transcribed in the form
\ba
\label{eq:ds-ham}
\hat H_\text{DS} & = & \rho_1 \hat n_d
+ \rho_2 \hat n_d(\hat n_d - 1) 
+ \rho_3 [-\hat C_{\mathrm{SO}(5)}
  + \hat n_d(\hat n_d + 3)] \nonumber\\*
&& + \rho_4 [\hat C_{\mathrm{SO(3)}} - 6\hat n_d] ~,
\label{H-DS}
\ea
where $\hat{C}_{\rm G}$ denotes a Casimir operator of
G, and 
$\hat{n}_d\!=\!\sum_{m}d^{\dag}_md_m\!=\!\hat{C}_{{\rm U(5)}}$. 
$\hat{H}_{\rm DS}$ is completely 
solvable
with eigenstates
$\ket{[N],n_d,\tau,n_{\Delta},L}$ and energies
$E_\text{DS} = \rho_1 n_d + \rho_2 n_d(n_d-1) 
+ \rho_3(n_d - \tau)(n_d + \tau + 3)
+ \rho_4 [L(L+1) - 6n_d]$.
The U(5)-DS spectrum is that of a spherical vibrator
with states arranged in
$n_d$-multiplets, the lowest ones being ($n_d\!=\!0,L\!=\!0$), 
($n_d\!=\!1,L\!=\!2$), ($n_d\!=\!2,L=4,2,0$),
($n_d\!=\!3,L\!=\!6,4,3,0,2$)
at energies $E(n_d)\approx n_d\,E(n_d\!=\!1)$.
The $E2$ operator in the IBM is proportional to
\begin{equation}
\label{Qchi}
\hat Q_\chi = d^\dagger s + s^\dagger \tilde d
+ \chi (d^\dagger \tilde d)^{(2)}~,
\end{equation}
where $\tilde d_m \!=\! (-1)^m d_{-m}$.
It is customary in the U(5)-DS limit to set $\chi=0$,
which results in vanishing quadrupole moments and
strong ($n_d+1\!\to\! n_d$) $E2$ transitions 
with particular ratios, {\it e.g.}, 
$\frac{B(E2;\,n_d+1,L'=2n_d+2\to n_d,L=2n_d)}
{B(E2;\,n_d=1,L=2\rightarrow n_d=0,L=0)}
=(n_d+1)\frac{(N-1)}{N}$.

The empirical spectrum of $^{112}$Cd, shown in
Fig.~\ref{fig:112cd-schematic}, consists
of both normal and intruder levels, the latter 
based on 2p-4h proton excitations across the
$Z\!=\!50$ shell gap.
At first sight, the normal states seem to 
follow the expected pattern of spherical-vibrator
$n_d$-multiplets. The measured $E2$ rates
support this view for the majority
of normal states, however, selected non-yrast states
(shown in red in Fig.~\ref{fig:112cd-schematic})
reveal marked deviations from this behavior.
Specifically, the $0^{+}_3$ and $2^{+}_4$ states
in $^{112}$Cd (denoted in Table~\ref{Tab-1} by
$0^{+}_{\alpha}$ and $2^{+}_{\alpha}$)
which in the U(5)-DS classification are members of the
$n_d=2$ and $n_d=3$ multiplets, respectively,
have unusually small $E2$ rates for the transitions
$0^{+}_{\alpha}\to 2^{+}_{1}$ and
$2^{+}_{\alpha}\to 2^{+}_{2}$, and large rates for
$0^{+}_{\alpha}\to 2^{+}_{2}$,
at variance with the U(5)-DS predictions.
Absolute $B(E2)$ values for transitions from the
$0^{+}_4$ state are not known, but its branching ratio to
the $2^{+}_2$ state is small.
As shown in Table~\ref{Tab-1},
the same unexpected decay patterns occur
in all $^{110-116}$Cd isotopes and comprise the so called
``Cd problem''~\cite{HeydeWood11}.
We are thus confronted with a situation in
which some states in the spectrum obey the predictions
of U(5)-DS, while other states do not. 
These empirical findings suggest the presence of a PDS,
as demonstrated for $^{110}$Cd in~\cite{Leviatan18}.
In what follows, we show that the same U(5)-PDS approach
is relevant also to the other Cd isotopes.
\begin{figure}[t]
\includegraphics[width=1\linewidth]{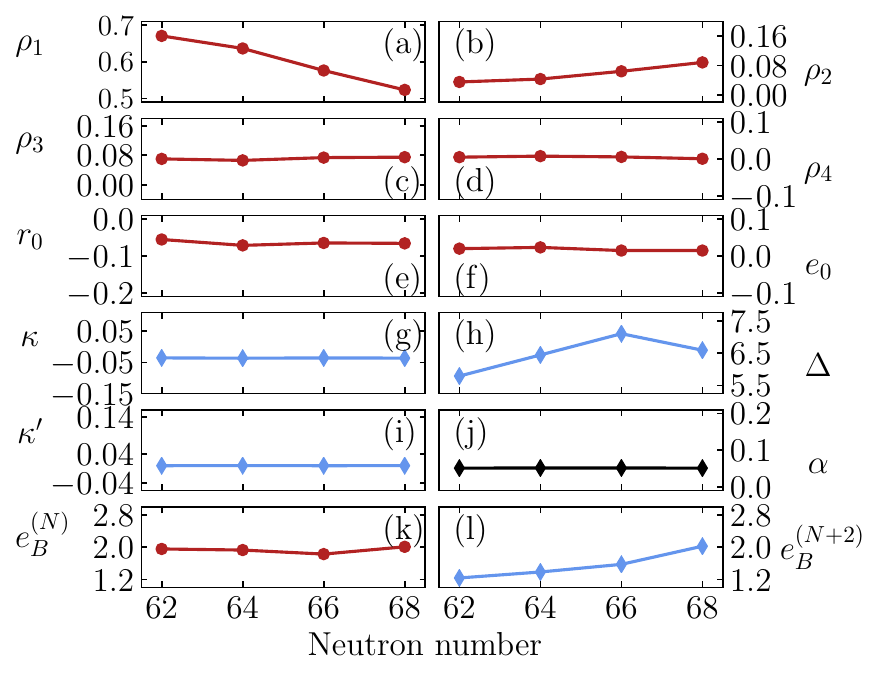}
\caption{Parameters of the IBM-CM Hamiltonian,
  Eq.~(\ref{Hfull}),
  in MeV and of the $E2$ operator, Eq.~(\ref{T-E2}),
  with $e^{(N)}_B$, $e^{(N+2)}_B$ in
  $\sqrt{\text{W.u.}}$ and $\chi_n \!=\! -0.7$,
  $\chi \!=\! -0.09$ are dimensionless.
  The boson numbers in the (normal, intruder) configurations
  are \mbox{$(N,N\!+\!2)$} with with $N=7,8,9$ and
  $\bar{N}=8$ (hole bosons)
for neutron numbers 62, 64, 66, and 68, respectively.
\label{fig:params}}
\end{figure}
\begin{figure}[t]
\begin{overpic}[width=1\linewidth]{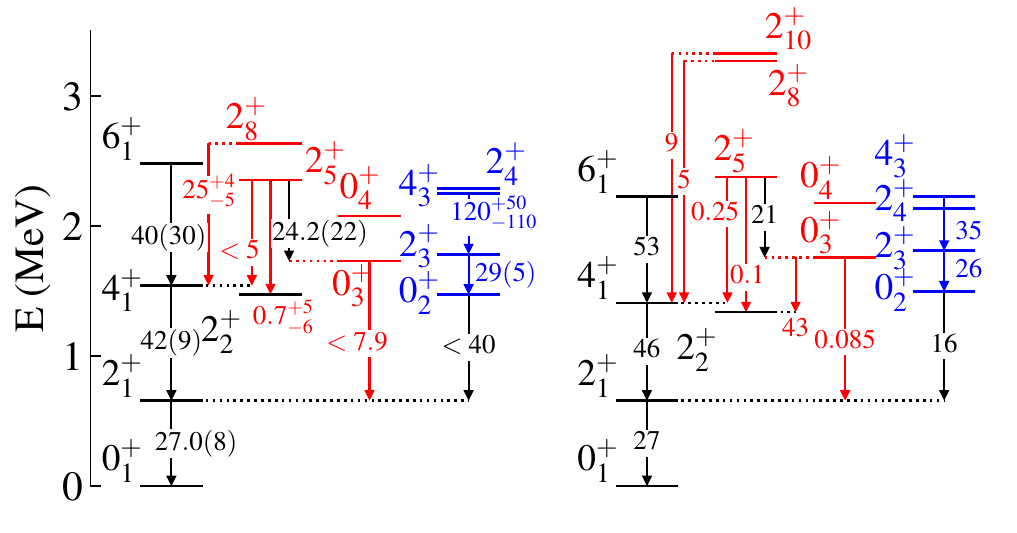}
\put (45,40) {\large $^{110}$Cd}
\put (10,45) {\large (a)}
\end{overpic}
\begin{overpic}[width=1\linewidth]{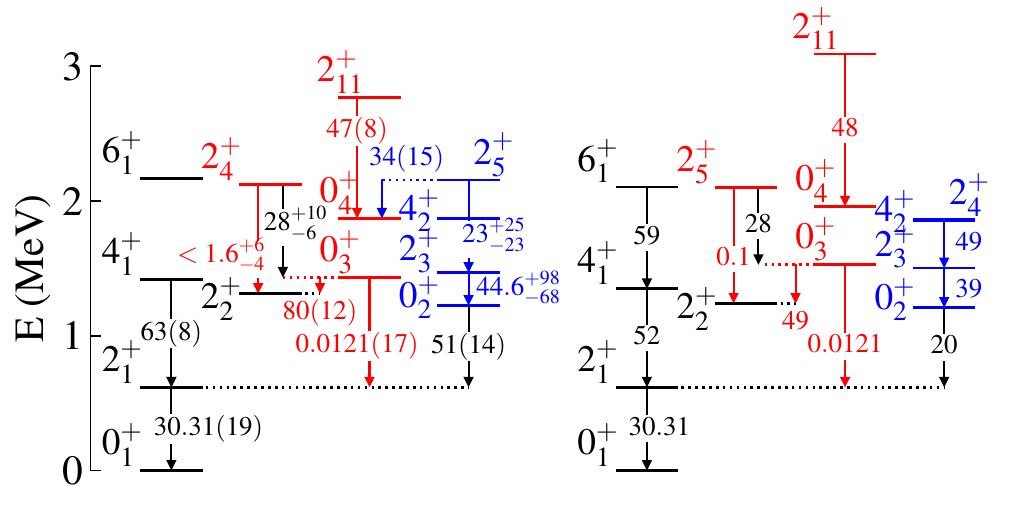}
\put (45,40) {\large $^{112}$Cd}
\put (10,40) {\large (b)}
\end{overpic}
\begin{overpic}[width=1\linewidth]{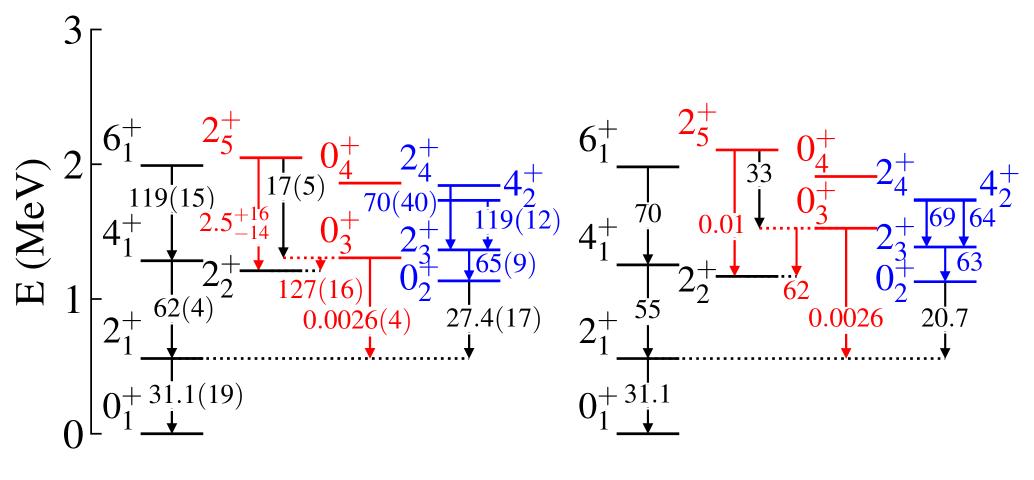}
\put (45,40) {\large $^{114}$Cd}
\put (10,40) {\large (c)}
\end{overpic}
\begin{overpic}[width=1\linewidth]{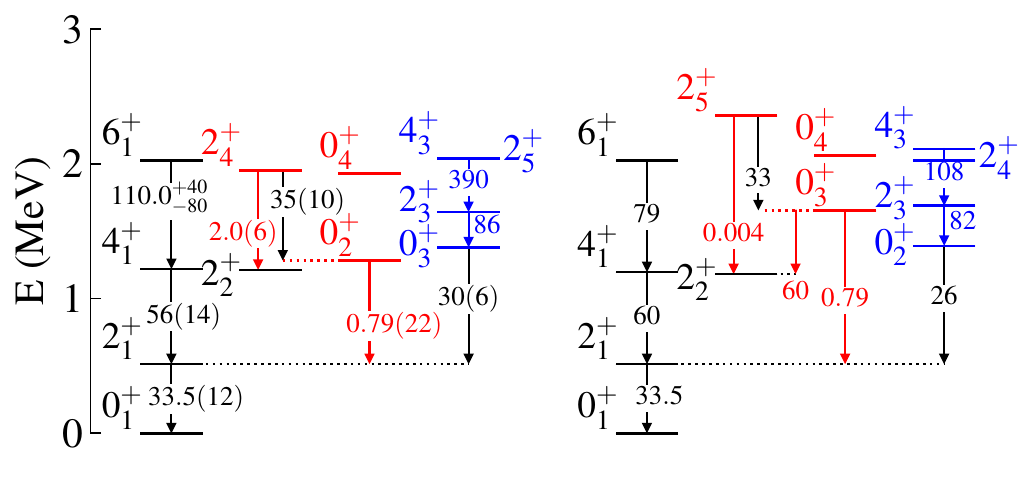}
\put (45,40) {\large $^{116}$Cd}
\put (10,40) {\large (d)}
\put (30,0) {\large EXP}
\put (70,0) {\large CALC}
\end{overpic}
\caption{Comparison between selected experimental
(left panels) and
  calculated (right panels) energy levels in MeV and
  $E2$ transition rates in W.u.
\label{fig:Cd-A}}
\end{figure}
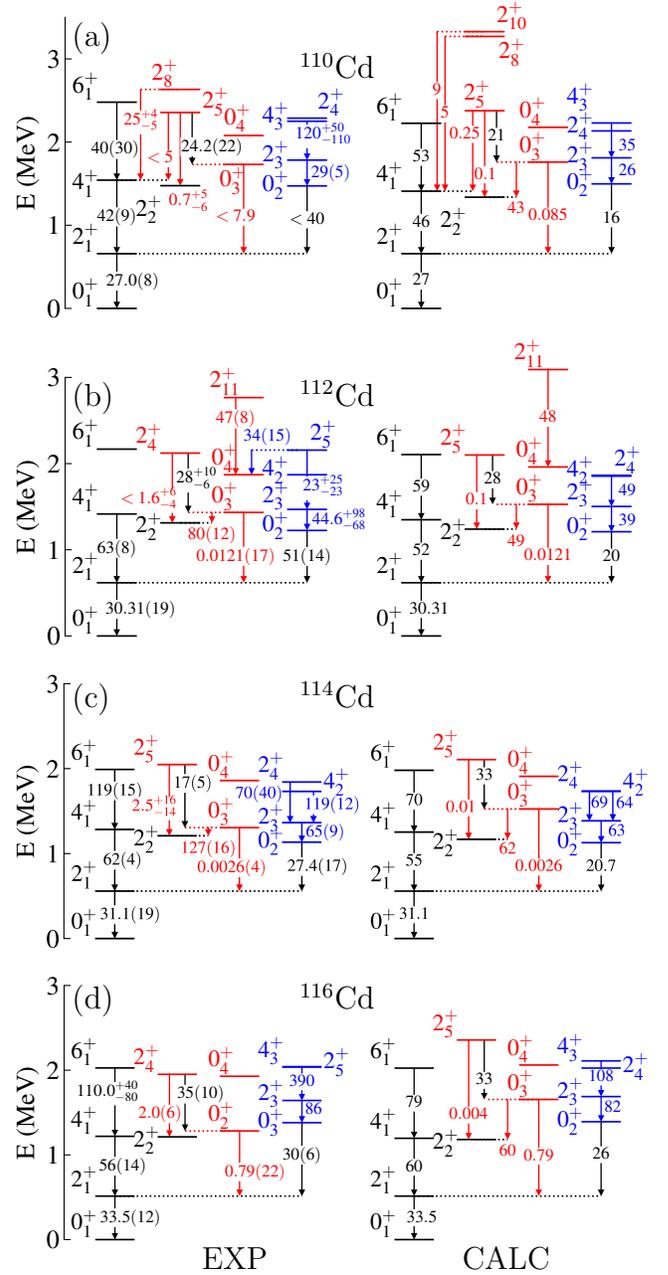

To describe both normal and intruder states,
we adopt the
interacting boson model with configuration mixing
(IBM-CM)~\cite{Duval81,Duval82},
widely used to study shape coexistence in
nuclei~\cite{Ramos14,Ramos15,Nomura18,Gavrielov22,Ramos22}.
The Hamiltonian is written as
\ba
\hat{H} = \hat{H}_{\rm PDS}^{(N)} + \hat{H}_{\rm intrud}^{(N+2)} 
+ \hat {V}_{\rm mix}^{(N,N+2)} ~,
\label{Hfull}
\ea
where the superscript $(N)$ denotes
a projection onto a space of $N$ bosons.
Here $\hat{H}_{\rm PDS}^{(N)}$ represents the normal
configuration ($N$ boson space),
$\hat{H}_{\rm intrud}^{(N+2)}$ represents
the intruder configuration ($N\!+\!2$ boson space)
and $\hat {V}_{\rm mix}^{(N,N+2)}$ a mixing term.
Explicit forms are given by
\bsub
\ba
\hat{H}_{\rm PDS} &=& \hat{H}_{\rm DS} +
r_0\, G^{\dag}_{0}G_{0}
+ e_{0}\, (G^{\dag}_0 K_0 + K^{\dag}_{0}G_0 ) ~,\quad
\label{H-PDS}\\
  \hat H_{\text{intrud}} &=&
  \kappa\hat{Q}_\chi\cdot\hat{Q}_\chi
  + \kappa^\prime\hat L\cdot\hat L + \Delta ~,
  \label{H-intrud}\\
\hat V_{\rm mix} &=&  
\alpha \left [(s^{\dagger})^{2}
  + (d^{\dagger}d^{\dagger})^{(0)}\right ] 
+ {\rm H.c.} ~,
\label{Vmix}
\ea
\esub
where $\hat{H}_{\rm DS}$ is the U(5)-DS Hamiltonian of
Eq.~(\ref{H-DS}), $\textstyle{G^{\dag}_{0} \!=\!
[(d^\dag d^\dag)^{(2)} d^\dag]^{(0)}}$, 
$K^{\dag}_{0} \!=\! s^{\dag}(d^{\dag} d^{\dag})^{(0)}$,
$\hat{Q}_{\chi}$ is given in Eq.~(\ref{Qchi})
and H.c. means Hermitian conjugate.
As shown in~\cite{Leviatan18}, $\hat{H}_{\rm PDS}$
has U(5)-PDS in the sense that it
breaks the U(5) symmetry, yet maintains
a subset of U(5)-DS basis
states $\ket{n_d=\tau, \tau, n_{\Delta}=0, L}$
with $L\!=\!\tau,\tau+1,\ldots,2\tau-2,2\tau$,
as solvable eigenstates. Henceforth, we refer to this
special subset of states as class-A states.
\begin{table*}
\begin{center}
  \caption{\label{Tab-2}
    \small
Normal-intruder mixing and U(5) structure of
the wavefunctions $\ket{\Psi,L}$, Eq.~(\ref{Psi}),
of selected eigenstates of $\hat{H}$, Eq.~(\ref{Hfull}).
Shown are the probability ($a^2$) of the normal part
$\Psi_n$, the dominant $n_d$ component in $\Psi_n$
and its probability $P_{n_d}$.}
\begin{tabular}{ccccclccclccclccc}
\hline\noalign{\smallskip}
  &  &
\multicolumn{3}{c}{$^{110}$Cd} & & 
\multicolumn{3}{c}{$^{112}$Cd} & & 
\multicolumn{3}{c}{$^{114}$Cd} & & 
\multicolumn{3}{c}{$^{116}$Cd} \\[1mm]
\cline{3-5}
\cline{7-9}
\cline{11-13}
\cline{15-17}\noalign{\smallskip}
$L^+_{k}$ & &
$a^2\,$(\%) & [($n_d$) $P_{n_d}$(\%)] & & & 
$a^2\,$(\%) & [($n_d$) $P_{n_d}$(\%)] & & &
$a^2\, $(\%) & [($n_d$) $P_{n_d}$(\%)] & & &
$a^2\,$(\%) & [($n_d$) $P_{n_d}$(\%)] & \\[1pt]
\noalign{\smallskip}\hline
$0^+_{1}$ & &
98.23  & [(0)$\;$ 98.22 ] & & &
97.94  & [(0)$\;$ 97.92 ] & & &
97.98  & [(0)$\;$ 97.95 ] & & &
98.27  & [(0)$\;$ 98.25 ] & \\
$2^+_{1}$ & &
96.38  & [(1)$\;$ 96.36 ] & & &
95.10  & [(1)$\;$ 95.05 ] & & &
95.28  & [(1)$\;$ 95.22 ] & & &
96.84  & [(1)$\;$ 96.81 ] & \\
$4^+_{1}$ & &
90.73  & [(2)$\;$ 90.69 ] & & &
83.19  & [(2)$\;$ 83.03 ] & & &
83.05  & [(2)$\;$ 82.87 ] & & &
92.95  & [(2)$\;$ 92.91 ] & \\
$2^+_{2}$ & &
89.81  & [(2)$\;$ 89.74 ]  & & &
81.62  & [(2)$\;$ 81.28 ] & & &
78.77  & [(2)$\;$ 78.33 ] & & &
91.31  & [(2)$\;$ 91.25 ] & \\
$6^+_{1}$ & &
71.18  & [(3)$\;$ 71.09 ] & & &
42.92  & [(3)$\;$ 42.53 ] & & &
39.46  & [(3)$\;$ 38.98 ] & & &
79.34  & [(3)$\;$ 79.27 ] & \\
$0^+_{\alpha}$ & &
70.75  & [(3)$\;$ 70.46 ] & & &
71.13  & [(3)$\;$ 69.54 ] & & &
71.55  & [(3)$\;$ 70.79 ] & & &
74.34  & [(3)$\;$ 74.14 ] & \\
$2^+_{\alpha}$ & &
68.34  & [(4)$\;$ 66.07 ] & & &
65.89  & [(4)$\;$ 62.83 ] & & &
40.78  & [(4)$\;$ 40.13 ] & & &
55.68  & [(4)$\;$ 54.73 ] & \\
\noalign{\smallskip}\hline
\end{tabular}
\end{center}
\end{table*} 
The eigenstates $\ket{\Psi;L}$
of $\hat{H}$, Eq.~(\ref{Hfull}),
involve normal ($\Psi_n$) and intruder ($\Psi_i$)
components in the $[N]$ and $[N+2]$ boson spaces,
\ba
\ket{\Psi; L} = a\ket{\Psi_n; [N], L}
  + b\ket{\Psi_i; [N+2], L}~,
\label{Psi}
\ea
with $a^{2}+b^{2}\!=\!1$.
The $E2$ operator in the IBM-CM reads
\ba
\label{T-E2}
\hat{T}(E2) = e^{(N)}_B \, \hat Q^{(N)}_{\chi_n} +
e^{(N+2)}_B\, \hat{Q}^{(N+2)}_\chi~,
\ea
with boson effective charges $e^{(N)}_B$ and $e^{(N+2)}_B$.

The parameters of $\hat{H}$~(\ref{Hfull}) and
$\hat{T}(E2)$~(\ref{T-E2}) are determined by
a combined fit to the spectra and $E2$ transitions
for the normal states ($2^+_1,4^+_1,2^+_2,6^+_1$)
and ($0^{+}_{\alpha},2^{+}_{\alpha}$), and for the lowest
($0^{+},2^{+}$) intruder states in each isotope.
As shown in Fig.~\ref{fig:params},
the extracted parameters
are fairly constant and vary smoothly as a function of
neutron number. Notable exceptions are
$\rho_1$ whose decrease
reflects the lowering of the $2^{+}_1$ state,
and $\Delta$, which together with the
$\kappa$ term in $\hat{H}_{\rm intrud}$ controls the
lowering of the intruder levels towards mid-shell
(neutron number 66), where boson particles
are replaced by boson holes.

An IBM-CM calculation has been performed for
spectral properties of states in $^{110-116}$Cd,
with energies up to 4 MeV.
A detailed account will be given in
a forthcoming longer publication~\cite{Gavrielov23}.
Here we focus on the main features which are relevant
to the subject matter of this Letter, namely, the
vibrational interpretation and symmetry aspects
of these isotopes.
The assignment of states as normal or
intruder, is based on their measured
$E2$ decays when available, or on their calculated
probabilities, $a^2$ and $b^2$, in Eq.~(\ref{Psi}).

As shown in Fig.~\ref{fig:Cd-A} and Table~\ref{Tab-1},
the U(5)-PDS calculation of
spectra and $E2$ rates
provides a good description 
of the empirical data in $^{110-116}$Cd.
It yields the same $B(E2)$ values as those of U(5)-DS
for class-A states and  reproduces correctly
the $E2$ transitions involving
the $(0^{+}_{\alpha},2^{+}_{\alpha}$) states
which deviate considerably from the U(5)-DS predictions. 
The origin of these features is revealed from
Table~\ref{Tab-2}, which shows for
eigenfunctions of $\hat{H}$ (\ref{Hfull}),
the percentage of the wave function
within the normal configuration [the probability $a^2$
of $\Psi_n$ in Eq.~(\ref{Psi})] and the dominant $n_d$
component in $\Psi_n$ and its probability.

The class-A states are dominated by the normal component
$\Psi_n$ (large $a^2\!\geq\! 90\%$), implying
a weak mixing (small $b^2$) with the intruder states.
The $6^{+}_1$ state experiences a larger mixing
consistent with its enhanced decay to the 
lowest $4^{+}$ intruder state. 
The $(0^{+}_{\alpha},2^{+}_{\alpha}$) states are more susceptible
to such mixing but still retain the dominance of $\Psi_n$ 
($a^2\!\sim\! 70\%$). For both types of states
the normal-intruder mixing increases with $L$ for a given
isotope, and increases towards mid-shell ($^{114}$Cd),
correlated with the decrease in energy of intruder states.
  
The class-A states possess
good U(5) quantum numbers to a good
approximation.
Their $\Psi_n$ part involves
a single $n_d$ component with probability
$P_{n_d}\geq 90\%$, as in U(5)-DS.
In contrast, the structure of the non-yrast
$(0^{+}_{\alpha},2^{+}_{\alpha}$) states
changes dramatically.
Specifically, the $\Psi_n$ parts of the $0^{+}_{\alpha}$
and $2^{+}_{\alpha}$ states, which in the U(5)-DS
classification have $n_d=2$ and $n_d=3$, have now dominant
components with $n_d=3$ and $n_d=4$, respectively.
The change $n_d\mapsto (n_d+1)$ ensures
weak ($\Delta n_d=2$) transitions from these states
to class-A states, but secures strong
$2^{+}_{\alpha}\to 0^{+}_{\alpha}$ ($\Delta n_d=1$) transitions,
in agreement with the data.
While the class-A and $(0^{+}_{\alpha},2^{+}_{\alpha}$) states
are predominantly spherical, the intruder states
are members of a single deformed band with a characteristic
$\gamma$-soft spectrum, shown in Fig.~\ref{fig:Cd-A},
and wavefunctions exhibiting
a broad $n_d$-distribution and a pronounced SO(6)
symmetry $\sigma=N\!+\!2$.

The PDS-CM describes the data very well, but there
are a few exceptions and remaining concerns.
The observed quadrupole moments $Q(2^{+}_1)$
and $B(E2;2^{+}_2\to 0^{+}_1)$,
shown in Table~\ref{Tab-1}, are larger than the
predicted values which, in turn, depend sensitively
on the choice of $\chi_n$ in Eq.~(\ref{T-E2}).
Larger values for these observables (which
involve class-A states) can be accommodated by adding
U(5) symmetry-breaking terms to the Hamiltonian.
The $(0^{+}_{\alpha},2^{+}_{\alpha}$) states
are predominantly $n_d=(3,4)$.
A relevant question~\cite{Garrett20},
is where their partner states with $n_d=(2,3)$ are
with enhanced decays to states with $n_d=(1,2)$.
The observed $0^{+}_4$ state, shown in Fig.~\ref{fig:Cd-A},
has a dominant branching to the intruder $2^{+}_{3}$ state,
hence does not match the properties expected for
a $n_d=2$ state.
This may indicate 
a different structure for the $0^{+}_4$ state ({\it e.g.}, a 4p-6h 
proton excitation as speculated in~\cite{Garrett12}), 
although fragmentation of $E2$ strength
cannot be ruled out. 
In $^{110}$Cd, the state  $2^{+}_8(2633)$ has a large 
$B(E2;  2^{+}_8\to 4^{+}_1)= 25^{+4}_{-5}$ W.u.~\cite{ensdf},
as expected for a $(n_d=3)\to (n_d=2)$ transition.
More data is needed to shed light on this issue.

The vibrational interpretation proposed here
is at variance with the microscopic BMF calculation of
Refs.~\cite{Garrett19,Garrett20} advocating multiple shape
coexistence in $^{110,112}$Cd, with states arranged in
deformed rotational bands.
Specifically, the states
$0^{+}_1,2^{+}_2,0^{+}_3,0^{+}_4$, of $^{112}$Cd,
shown in Fig.~\ref{fig:112cd-schematic}, 
serve as bandheads for
the ground, $\gamma$ and two excited $K\!=\!0$ bands,
and $0^{+}_2,2^{+}_5$ are bandheads for intruder and
intruder-$\gamma$ bands. Similar assignments were suggested
for $^{110}$Cd. The BMF-based approach is
parameter-free and provides
a qualitative description of $^{110,112}$Cd, but
with noticeable shortcomings.
In particular, the predicted energies
are generally overestimated, and in-band $B(E2)$ values
and quadrupole moments are greater than observed,
reflecting too large deformations in the calculated states.
A detailed comparison between the
BMF-based approach and the current PDS-based approach
will be given in~\cite{Gavrielov23}, with a view that
ultimately, comparison with data should be the basis to
accept or refute a model. One possible signature that can
distinguish between the two approaches is to measure the
value of $B(E2;4^{+}_2\to 3^{+}_1)$, which is expected to be
small (large) in the PDS (BMF) approach, where the
indicated states are in the same $n_d$-multiplet
(in the same $\gamma$ band).

To summarize, consistent with the empirical data,
we have shown that a vibrational
interpretation and good U(5) symmetry
are maintained for the majority of low-lying
normal states, coexisting with a single deformed band
of intruder states in $^{110,112,114,116}$Cd isotopes.
The observed deviations from this
paradigm, are properly treated by an Hamiltonian
which breaks the U(5) symmetry in selected non-yrast
states, while keeping the mixing with intruder states weak.
The results demonstrate, for the first time, the relevance
of a partial dynamical symmetry (PDS)
to a series of isotopes, and set the path for implementing
a similar PDS-based approach to other regions of the
nuclear chart, where a prescribed collective
structure paradigm holds for a segment of the spectrum.

This work was supported, in part, (A.L. and N.G.)
by the Israel Science Foundation and (J.E.G.R.) by
project PID2019-104002GB-C21 funded by
MCIN/AEI/10.13039/50110001103 and ``ERDF A way of
making Europe''.


\begin{thebibliography}{99}

\bibitem{Bohr36}
N.~Bohr,
Nature \textbf{137} 344, (1936).

\bibitem{Bohr37}
N.~Bohr and F.~Kalckar,
Mat.\ Fys.\ Medd.\ Dan.\ Vid.\ Selsk.\ \textbf{14} no~10, (1937).

\bibitem{Rainwater50}
J.~Rainwater,
Phys.\ Rev.\ \textbf{79} 432, (1950).

\bibitem{Bohr52}
A.~Bohr,
Mat.\ Fys.\ Medd.\ Dan.\ Vid.\ Selsk.\ \textbf{26} no~14, (1952).

\bibitem{Bohr53}
A.~Bohr and B.~R.~Mottelson,
Mat.\ Fys.\ Medd.\ Dan.\ Vid.\ Selsk.\ \textbf{27} no~16, (1953).

\bibitem{Bohr75}
A.~Bohr and B.~R.~Mottelson,
{\it Nuclear Structure. II Nuclear Deformations}
(Benjamin, New York, 1975).

\bibitem{Iachello87}
F.~Iachello and A.~Arima,
{\it The Interacting Boson Model}
(Cambridge University Press, Cambridge, 1987).

\bibitem{Garrett22}
  P. E. Garrett, M. Zieli\'nska, E. Cl\'ement,
  Prog. Part. Nucl. Phys. \textbf{124}, 103931 (2022).

\bibitem{Casten00}
R.~F.~Casten,
{\it Nuclear Structure from a Simple Perspective}
(Oxford University Press, Oxford, 2000).

\bibitem{Heyde04}
K.~Heyde,
{\it Basic Ideas and Concepts in Nuclear Physics}
(Institute of Physics, Bristol, 2004).

\bibitem{Rowe10}
D.~J.~Rowe and J.~L.~Wood,
{\it Fundamentals of Nuclear Models}
(World Scientific, Singapore, 2010).

\bibitem{Garrett07}
P.~E.~Garrett, K.~L.~Green, H.~Lehmann, J.~Jolie, C.~A.~McGrath,
M.~Yeh and S.~W.~Yates,
Phys. Rev. C \textbf{75}, 054310 (2007).

\bibitem{Casten92}
R.~F.~Casten, J.~Jolie, H.~G.~B\"orner, D.~S.~Brenner, N.~V.~Zamfir,
W.-T.~Chou and A.~Aprahamian, 
Phys. Lett. B \textbf{297}, 19 (1992).

\bibitem{Kadi03}
M.~Kadi, N.~Warr, P.~E.~Garrett, J.~Jolie and S.~W.~Yates,
Phys. Rev. C \textbf{68}, 031306(R) (2003).

\bibitem{Garrett08}
P.~E.~Garrett,  K.~L.~Green and J.~L.~Wood,
Phys. Rev. C \textbf{78}, 044307 (2008).

\bibitem{Garrett10}
P.~E.~Garrett and J.~L.~Wood,
J. Phys. G \textbf{37}, 064028 (2010); 069701 (2010).

\bibitem{Garrett12}
  P.~E.~Garrett, J.~Bangay, A.~Diaz Varela, G.~C.~Ball, D.~S.~Cross,
  G.~A. Demand {\it et al.},
Phys. Rev. C \textbf{86}, 044304 (2012).

\bibitem{Garrett19}
  P.~E.~Garrett, T.~R.~Rodr\'iguez, A.~D.~Varela, K.~L.~Green, J.~Bangay,
  A.~Finlay {\it et al.},
Phys. Rev. Lett. \textbf{123}, 142502 (2019).

\bibitem{Garrett20}
  P.~E.~Garrett, T.~R.~Rodr\'iguez, A.~Diaz Varela, K.~L.~Green, J.~Bangay,
  A.~Finlay {\it et al.},
Phys. Rev. C \textbf{101}, 044302 (2020).

\bibitem{Leviatan11}
A.~Leviatan,
Prog.\ Part.\ Nucl.\ Phys.\ \textbf{66}, 93 (2011).

\bibitem{Leviatan18}
A.~Leviatan, N.~Gavrielov, J.~E.~Garc\'\i a-Ramos, and P.~Van~Isacker,
Phys.\ Rev.\ C \textbf{98}, 031302(R) (2018).

\bibitem{Gneuss71}
G.~Gneuss and W.~Greiner,
Nucl.\ Phys.\ A \textbf{71}, 449 (1971).

\bibitem{Fielding77}
H.~W.~Fielding {\it et al.},
Nucl. Phys. A \textbf{281}, 389 (1977).

\bibitem{Kumpulainen92}
J.~Kumpulainen, R.~Julin, J.~Kantele, A.~Passoja, W.~H.~Trzaska, 
E.~Verho, J.V\"a\"ar\"am\"aki, D.~Cutoiu, and M.~Ivascu,
Phys.\ Rev.\ C \textbf{45}, 640 (1992).

\bibitem{HeydeWood11}
K. Heyde and J. L. Wood,
Rev. Mod. Phys. \textbf{83}, 1467 (2011).

\bibitem{Deleze93}
M.~D\'el\`eze, S.~Drissi, J.~Kern, P.~A.~Ternier, J.~P.~Vorlet,
J.~Rikovska, T.~Otsuka, S.~Judge, and A.~Williams,
Nucl.\ Phys.\ A \textbf{551}, 269 (1993).

\bibitem{Heyde82}
K.~Heyde, P.~Van~Isacker, M.~Waroquier, G.~Wenes, and M.~Sambataro,
Phys.\ Rev.\ C \textbf{25}, 3160 (1982).

\bibitem{ensdf}
Evaluated Nuclear Structure Data File (ENSDF),
\href{https://www.nndc.bnl.gov/ensdf/}{https://www.nndc.bnl.gov/ensdf/}.

\bibitem{Duval81}
P.~D. Duval and B.~R. Barrett, 
Phys. Lett. B \textbf{100}, 223 (1981).

\bibitem{Duval82}
P. D. Duval and B. R. Barrett,
Nucl. Phys. A \textbf{376}, 213 (1982).

\bibitem{Ramos14}
J.~E.~Garc\'\i a-Ramos and K.~Heyde,
Phys. Rev. C \textbf{89}, 014306 (2014).

\bibitem{Ramos15}
J.~E.~Garc\'\i a-Ramos and K.~Heyde,
Phys. Rev. C \textbf{92}, 034309 (2015).

\bibitem{Nomura18}
  K.~Nomura and J.~Jolie,
  Phys. Rev. C {\bf 98}, 024303 (2018).

\bibitem{Gavrielov22}
N. Gavrielov, A. Leviatan and F. Iachello,  
Phys. Rev. C \textbf{105}, 014305 (2022).

\bibitem{Ramos22}
E. Maya-Barbecho and J.~E.~Garc\'\i a-Ramos,
Phys. Rev.~C \textbf{105}, 034341 (2022).

\bibitem{Gavrielov23}
  N. Gavrielov, P. Van Isacker, J.~E.~Garc\'\i a-Ramos
  and A.~Leviatan, (to be published).


\end{thebibliography}
\end{document}